# The “Bayesian” brain, with a bit less Bayes

Eelke Spaak, eelke.spaak@donders.ru.nl

Donders Institute for Brain, Cognition, and Behaviour
Radboud University
Nijmegen, The Netherlands

---

**Abstract**

The idea that the brain is a probabilistic (Bayesian) inference machine, continuously trying to figure out the hidden causes of its inputs, has become very influential in cognitive (neuro)science over recent decades. Here I present a relatively straightforward generalization of this idea: the primary computational task that the brain is faced with is to track the probabilistic structure of observations themselves, without recourse to hidden states. Taking this starting point seriously turns out to have considerable explanatory power, and several key ideas are developed from it: (1) past experience, encoded in prior expectations, has an influence over the future that is analogous to regularization as known from machine learning; (2) action generation (interpreted as constraint satisfaction) is a special case of such regularization; (3) the concept of attractors in dynamical systems provides a useful lens through which prior expectations, regularization, and action induction can be viewed; these thus appear as different perspectives on the same phenomenon; (4) the phylogenetically ancient imperative of acting to ensure and thereby observe conditions beneficial for survival is likely the same as that which underlies perceptual inference. The Bayesian brain hypothesis has been touted as promising to deliver a “unified science of mind and action”. In this paper, I sketch an informal step towards fulfilling that promise, while avoiding some pitfalls that other such attempts have fallen prey to.

---

## 1. Introduction

In recent decades, the idea of the “Bayesian brain” has become very influential in cognitive neuroscience and related disciplines. It appears in various guises, from predictive processing to the free-energy principle; and the general gist in most instances is typically the following. Organisms are only ever presented with noisy sensory observations, which are by themselves not sufficient to reach any conclusions about the structure of the world that elicited them. Therefore, organisms have to rely on prior information about the structure of this world in order to make inferences. Specifically, the primary computational task of any organism’s brain is to uncover the latent causes that gave rise to its observations. Phrased probabilistically, the brain is trying to infer p(s|o), the probability of (unobserved) world state *s*, after observing *o*. What is the likely state of the world, given my current sensory input?

As is well known through Bayes' famous theorem, this posterior probability p(s|o) is proportional to the product of the probability p(o|s) (often treated as a "likelihood" function, describing how a given world state should result in some particular observation) and the prior probability over world states p(s). Solving the problem of inferring p(s|o) requires Bayesian inference, hence the name of this now prominent neuroscientific paradigm. There is a wealth of empirical evidence that supports the idea that cognition and the brain indeed function largely according to Bayesian principles (Clark, 2013; de Lange et al., 2018; Doya et al., 2006; Friston, 2010; Rao & Ballard, 1999).

Several open questions about the brain's Bayesian nature remain, however. The ultimate task of the nervous system is to control the body. This fact appears at odds with many formulations of the Bayesian brain hypothesis: if the nervous system has *one* primary task, then which is it: inferring hidden states or controlling the body? Might we be able to cast the one in terms of the other? Furthermore, in order to conduct Bayesian inference about hidden world states, one needs prior expectations. Where do these priors over world states, and likelihoods that generate new potential observations, come from? Might these, too, be a special consequence of a general principle?

In this paper, I develop several points related to the "Bayesian brain" hypothesis. As a starting point, I will emphasize how a common and typical interpretation of this idea – the brain needs to infer hidden states given sensory evidence – is a limited one. This is only a special case of a more general overarching goal; a special case that appears evident when one focuses on sensory systems. The underlying overarching goal for the nervous system is to continuously attempt to optimally match the distribution over observations themselves, without recourse to hidden states. In other words, the nervous system is attempting to match the so-called *marginal* distribution p(o), rather than the *conditional* distribution p(s|o). The only interface between the nervous system and the world is through the observations that the world imparts upon the nervous system (through the various senses), and the manipulations that the nervous system imparts upon the world through muscle tissue (in turn influencing observations).[1]

Therefore, as a natural starting point, all that exists for the nervous system to track is this marginal distribution p(o). I aim to demonstrate here that taking this viewpoint seriously has several key advantages. Specifically, a nervous system that optimally attempts to match p(o) will “automatically” generate behaviour, as outlined below. Furthermore, by taking this viewpoint to its logical conclusion, we will see that perceptual inference, efficient coding of sense data, and affecting the world through behaviour are all different perspectives on the same underlying phenomenon: attractor states in the input space of an agent. In this way, the narrative presented here offers an informal step towards fulfilling the promise of the Bayesian brain as a “unified science of mind and action” (Clark, 2013).

[1] Giving center stage to the marginal distribution over observations has predecessors in spirit (though using different terminology) in the work of early cyberneticists like Ashby, who famously wrote that "the whole function of the brain is summed up in: error correction" – a quote used by Andy Clark to open his influential paper 'Whatever next?' (Clark, 2013). Many authors who may embrace this starting point, including Clark, tend to subsequently and immediately make the move to a distinctively Helmholtzian view of what this entails: 'sensory systems are in the tricky business of inferring sensory causes from their bodily effects' (Clark, 2013; Helmholtz, 1962). Part of my goal here is to demonstrate how this move in emphasis detracts from the explanatory power of the overarching principle. (See also (Bruineberg et al., 2018) for a critique of the dominance of this Helmholtzian interpretation of the Bayesian brain hypothesis on enactivist grounds.)

One key way of reading this paper is as primarily *didactic* in nature. Although I develop some theoretical arguments, a main goal is to (I) provide a clear outline of several core ideas and problems in the contemporary narrative on the ‘Bayesian brain’, while (II) highlighting how various concepts of diverse heritage, already well-established in isolation, can fruitfully come together, and in doing so perhaps enrich the reader’s conceptual thought. To that end, the paper is written as much as possible without complicated jargon, and with reference to sometimes ‘nichey’ discussions kept to a minimum as well. Because of this, the paper should be accessible to a broad audience of cognitive (neuro)scientists. However, this also entails that some concepts many readers will be familiar with are explained in more detail than strictly necessary, while other concepts are not given as thorough a treatment as might be fitting. I nonetheless hope that the overall narrative proves insightful to most, and that even those who may not agree with its main conclusions take something away from the road that led there.

Existing work on active inference and the free energy principle – to which the present work is indebted – has previously proposed an intimate link between action generation and perceptual inference (Friston, 2010; Friston et al., 2015). However, in doing so, it has ended up conflating expectations and desires (Yon et al., 2020) and adopted technical terms like ‘Markov blankets’ using problematic definitions (Bruineberg et al., 2021), not rarely causing deep confusion (Freed, 2010). In the present work, I strive to avoid these issues, by describing the shared computational machinery underlying perception and action in novel terms. I focus on bringing together different strands of thought, previously disjoint, and leave details of implementation aside.

This paper is structured as follows. In section 2, I start from the viewpoint of probabilistic inference regarding the marginal distribution of observations, without reference to hidden states. I outline how the imperative for encoding observations, and extracting useful information from them, results in prior expectations influencing perception, which in turn have an effect akin to *regularization*. In section 3, I describe how such regularizing prior factors over observations, although introduced from the perceptual angle, can equivalently be viewed as inducers of action. In section 4, I describe the physical concept of attractor states in dynamical systems, and highlight how this concept may unify prior expectations, probabilistic sampling, regularization, and action generation. Finally, in section 5, I return to action generation and outline how the Bayesian machinery we now know is involved in perceptual systems turns out to have its roots in action.

## 2. Perceptual inference without hidden states

### 2.1 Tracking the probabilistic structure of observations is energy-efficient

Imagine a *tabula rasa* agent equipped with the ability to observe. This agent has no expectation whatsoever about what observations it might encounter. In other words, at the onset of its life, its expectation about p(o) is well modelled with a uniform distribution, as shown in Figure 1:

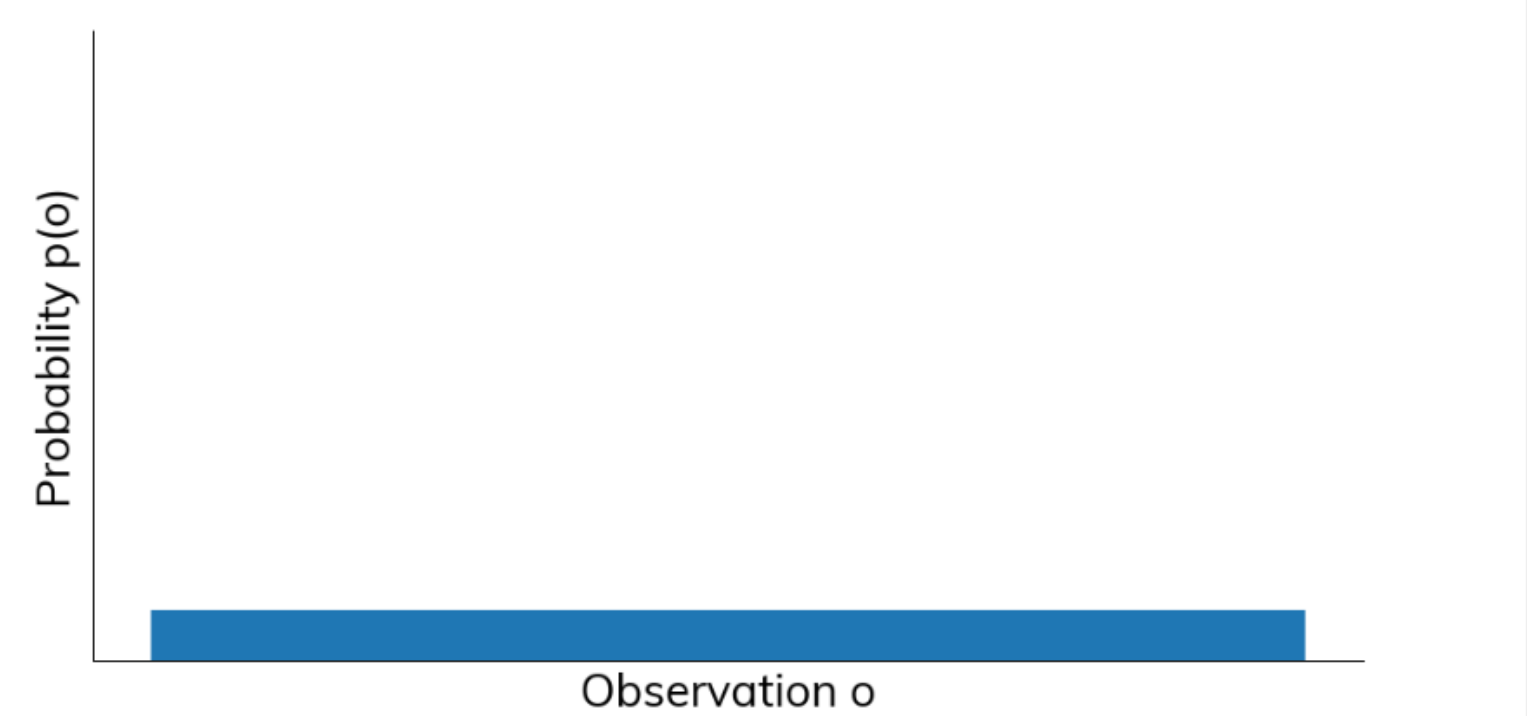

**Figure 1**. The uniform distribution: each possible observation is expected with equal probability.

For illustration, I have restricted this problem here to one-dimensional and discrete. One-dimensional is clearly biologically implausible, as organisms of interest typically have very many sensory cells. Discreteness appears a more plausible assumption, since the number of possible observations is probably finite, though it is certainly huge. Therefore, for now, I am additionally restricting the number of possible observations to 1,024.

Whenever an agent encounters a particular observation, it will want to encode this somehow; it will want to use the information imparted by this observation and make it available to its internal circuitry for generating behaviour. Given the above expected distribution at the beginning of this agent's life, when it encounters its first ever observation, $o_1$, how could the agent encode this? An obvious way is to encode all possible observations by a unique identifier, something like: "$o_1$ is observation type 723 out of all possibilities". Equivalently, we can describe such an encoding scheme by saying the agent adopts a binary code: "$o_1$ = 1011010011". Since there are 1,024 possible observations, this encoding scheme requires $\log_2(1024) = 10$ bits per observation. This value is known as the (Shannon) entropy of this distribution, and the value of 10 is in fact the maximum possible entropy for any discrete distribution over 1,024 possible values (entropy is maximum for the uniform distribution).

The amount of resources (or energy) needed to encode observations under a particular distribution is directly related to the entropy of that distribution (Bérut et al., 2012; Landauer, 1961)[2]. That is, the fewer bits an agent requires to describe an observation, the more energy-efficient these descriptions will be. Although in this example, 10 bits seems very little, remember that this is an extreme simplification and the entropy of the uniform distribution over all possible sensory inputs across all sensory cells of an actual organism is actually massive. Such a coding scheme would require prohibitively large codes for each possible observation and thus require prohibitively large amounts of energy. Therefore, a more efficient scheme than coding all possible observations with unique identifiers is essential if the agent wants to do useful work based on incoming information.

After our agent has lived a small portion of its life (let's say 20 timesteps), it will have encountered 20 observations, as shown in Figure 2:

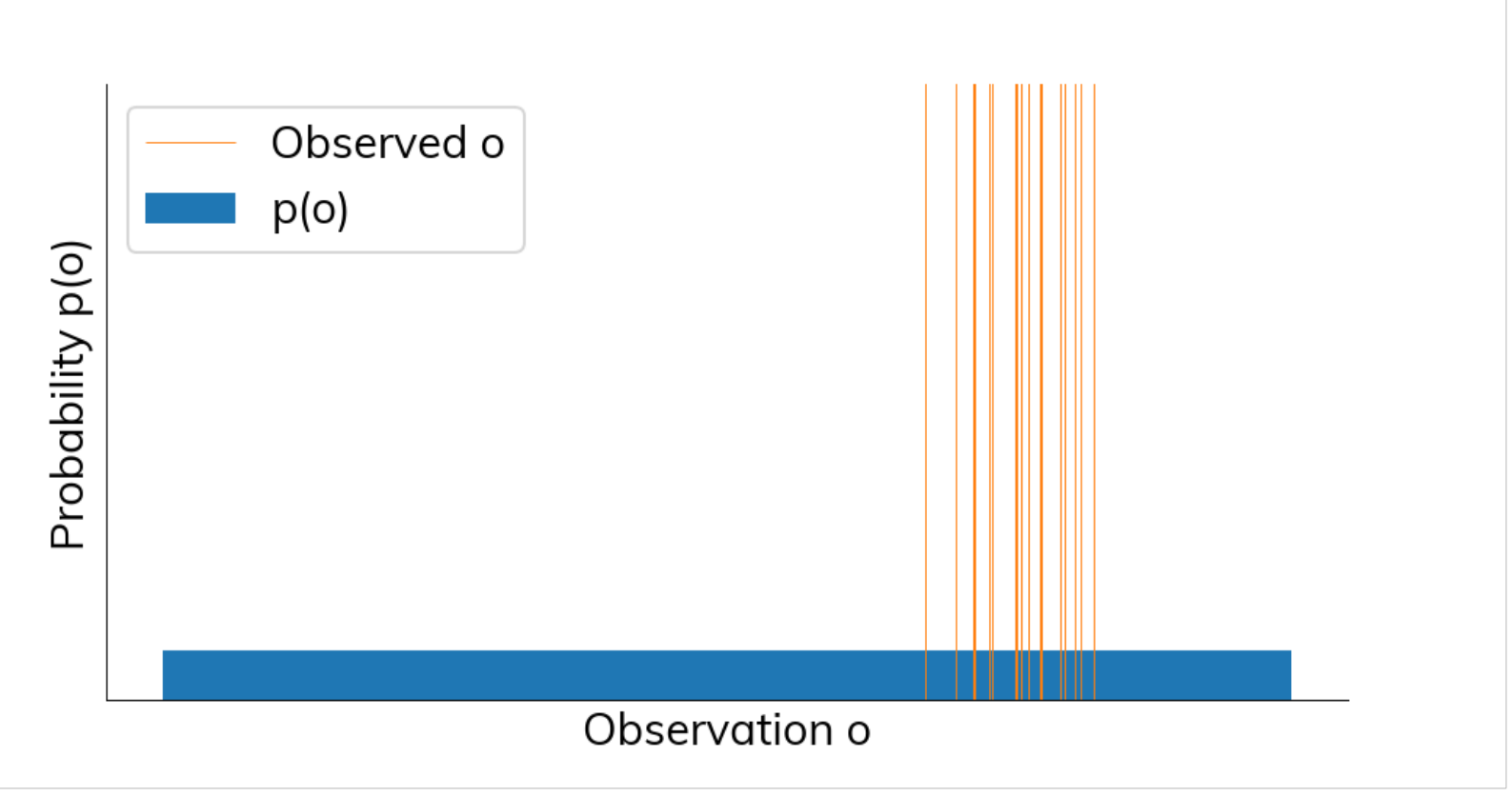

**Figure 2.** Actually encountered observations $o_1$ through $o_{20}$ (orange lines), superimposed upon the expected p(o).

Based on these 20 observations, it seems that perhaps the uniform distribution for p(o) is not a great description of the actual probabilistic structure of observations. Observations are coming in clustered much closer together than one would expect if the underlying process truly were uniform. Our agent might leverage this structure in order to more efficiently encode newly incoming observations. The only assumption needed for this to work is that the world that generates observations is similar between one time point and the next. While I'm positing this property here as an assumption, its truth can be appreciated from noting that any organisms of interest have evolved in our world over many millions of years. If the essential structure of the world had not remained relatively constant across such time scales, then evolution would not have had a chance to devise the intricate organisms we see today. Natural selection can only result in the *accumulation* of inherited traits in a reasonably stable world, thus organisms who owe their existence to natural selection will be equipped with systems well adapted to such stability.

[2] The references cited demonstrate the direct link between thermodynamic energy and information in physical systems, establishing a lower bound on the energy needed to encode or delete one bit. In actual organisms, the amount of energy needed to encode information is almost certainly much larger than this lower bound.

How can our agent use the world's stability in order to more efficiently encode its observations? By assigning higher probability to those regions of observation space that are similar to past observations (Figure 3):

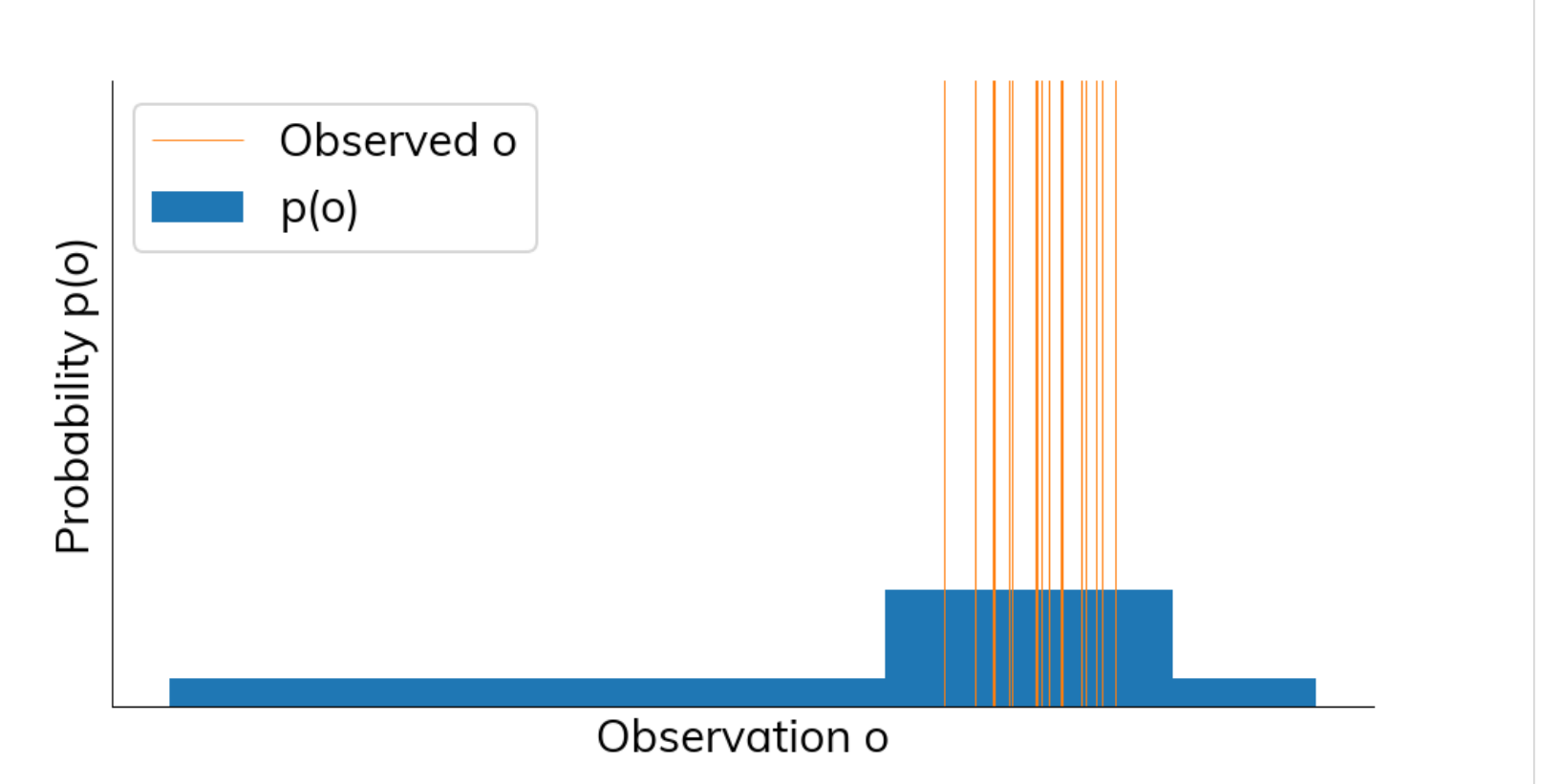

**Figure 3.** A perhaps better approximation (in blue) of the statistical structure of observations so far?

In this case, the agent has quadrupled the probability by which it expects possible observations 640 through 896. A very simple new and more efficient coding scheme is the following. If a new observation $o_{21}$ comes in, the agent could encode it by using the first bit to indicate whether $o_{21}$ comes from the high-density region or not. If so, then the remaining bits can be used to indicate which of the observations from this region was actually observed, and that can be done in $\log_2(256) = 8$ bits. In total, for the majority of possible observations, the agent then only requires a total of 9 bits, instead of 10. Of course, observations in the low-density region are still possible (their $p(o) > 0$), and they require now 11 bits to encode, but since they are much less likely, overall, this coding scheme saves on encoding space and thereby is more energy efficient. In fact, this simple coding scheme is not even close to the most efficient one that is possible, given the probability distribution just described. The average number of bits required to encode observations uniquely in this scheme using the most efficient encoding possible is again given by the entropy of the distribution, which in this case is around 6.7 bits. In other words, discovering that the probability distribution over possible observations is non-uniform (in the way depicted in Figure 3) has saved our agent an average of 34% of coding space (and, relatedly, energy) for describing them.[3]

## 2.2 Energy efficiency, compressibility, and information extraction

The way I described the efficiency of an information coding scheme in terms of the length of the bit string required for encoding one observation suggests a close link to *compressibility*.

[3] I have deliberately described the principle of efficient coding using a toy example. It is worth noting that both theoretical and empirical work exists that demonstrates that sensory neurons indeed adopt efficient coding principles, specifically by leveraging the statistics of their environment (Atick et al., 1992; Atick & Redlich, 1990; Barlow, 1961).

Shorter descriptions of pieces of information are compressed forms of that same information. (As an aside, it has been proposed that compressibility of incoming sensations by an observer, and the change in compressibility over time, are intimately linked to subjective experiences of beauty and interestingness, respectively (Schmidhuber, 2009).) Related to compressibility is the notion of Kolmogorov complexity (also known as algorithmic entropy), the length of the shortest possible algorithm that generates a piece of information. Short algorithms that produce some longer output X are compressed forms of X. If our agent learns better (more efficient) encoding schemes for incoming observations *o* by leveraging the structure of past p(o), it is in essence discovering algorithms that would generate such *o*. I am here primarily focusing on the Shannon (probabilistic) view of information for illustration, rather than the algorithmic one, because it is likely better known. However, the algorithmic interpretation of information actually has some properties that make it better suited for understanding information in biological systems; refer to (Grünwald & Vitányi, 2003) for insightful discussion about parallels and differences between the two senses of information.

So far I've described the primary rationale for the agent's striving to approximate p(o) in terms of the (energy) efficiency of the encoding scheme for incoming observations. Importantly, an efficient code for an observation *o* is also a form of information *extraction* from that observation. If we were to encounter the bit string 011011011011, we could describe this by saying something like "the first position is 0, the second position is 1, the third position is 1, the fourth position is 0, …, the twelfth position is 1", which is analogous to the uniform distribution coding scheme discussed before. However, an alternative description here would be "the first position is 0, the second position is 1, the third position is 1, and repeat this pattern four times". This is not only a more efficient description in terms of length and energy required, but in an important sense, contains more information than the raw observation itself. It allows for the faithful reconstruction of the full bit sequence, but additionally tells us something about its internal structure. And even if a description does *not* allow for a full reconstruction of the raw string, it can still offer information about that string that was not obvious without such description. (In other words, even 'lossy' compressions are valuable.) The sense in which such shorthand descriptions offer *more* information than the raw data is not well captured by the notion of Shannon information. A more useful perspective is perhaps Dennett's "real pattern" (Dennett, 1991).

A full discussion of Dennett's perspective is beyond the scope of this paper, but briefly: 'here is an elephant' is a much more efficient description of a certain state of affairs than 'here is a grey skin cell, here is another grey skin cell, etc.'. This is true on the Shannon information account: 'elephant' (abstracting away from noise) uses fewer bits than describing all the individual cells; as well as on the Dennettian account: 'elephant' leverages a real pattern among the noise, which is therefore efficient. However, the Shannon account primarily emphasizes that the 'elephant' shorthand contains less information (namely: fewer bits are needed) than the cell-level account. With Dennett (presumably), I would emphasize that the 'elephant' description (also) contains, in an important sense, *more* information than the cell-level account: "If one finds a predictive pattern of the sort just described one has ipso facto discovered a causal power – a difference in the world that makes a subsequent [testable] difference" (Dennett, 1991, p. 43-44, note 22). For this to be true, one does not need to make the additional assumption or inference that there apparently exists a 'hidden state' that may be the cause of such a pattern; the discovery of the pattern in the input is sufficient for both increased coding efficiency and the establishment of predictive power.

## 2.3 Tracking probabilistic structure by continuous model updates

Back to our agent. It can learn to leverage the probabilistic structure of incoming observations in order to construct an efficient coding scheme, which furthermore entails the extraction of *useful* information. It should be noted that the mathematically optimal way of updating estimates of probability distributions after observing some data is exactly given by Bayesian inference. The phrase "a bit less Bayes" in the title of this paper is not meant to signal that Bayesian inference has no role to play in such updates. Instead, it is meant to indicate that reducing the focus on *hidden states* (of the world) can be beneficial in understanding the nervous system.

The agent's approach of simply quadrupling the probability mass in a fixed region around its first twenty observations (as in Figure 3) was crude, yet at least somewhat effective. Primarily, this allowed me to describe relatively straightforwardly the relationship between probability and coding efficiency. However, in reality, the approximation of p(o) will likely not happen quite so abruptly or in such a "blocky" fashion. Instead, each observation might impart a bit of probability mass after being observed (Figure 4):

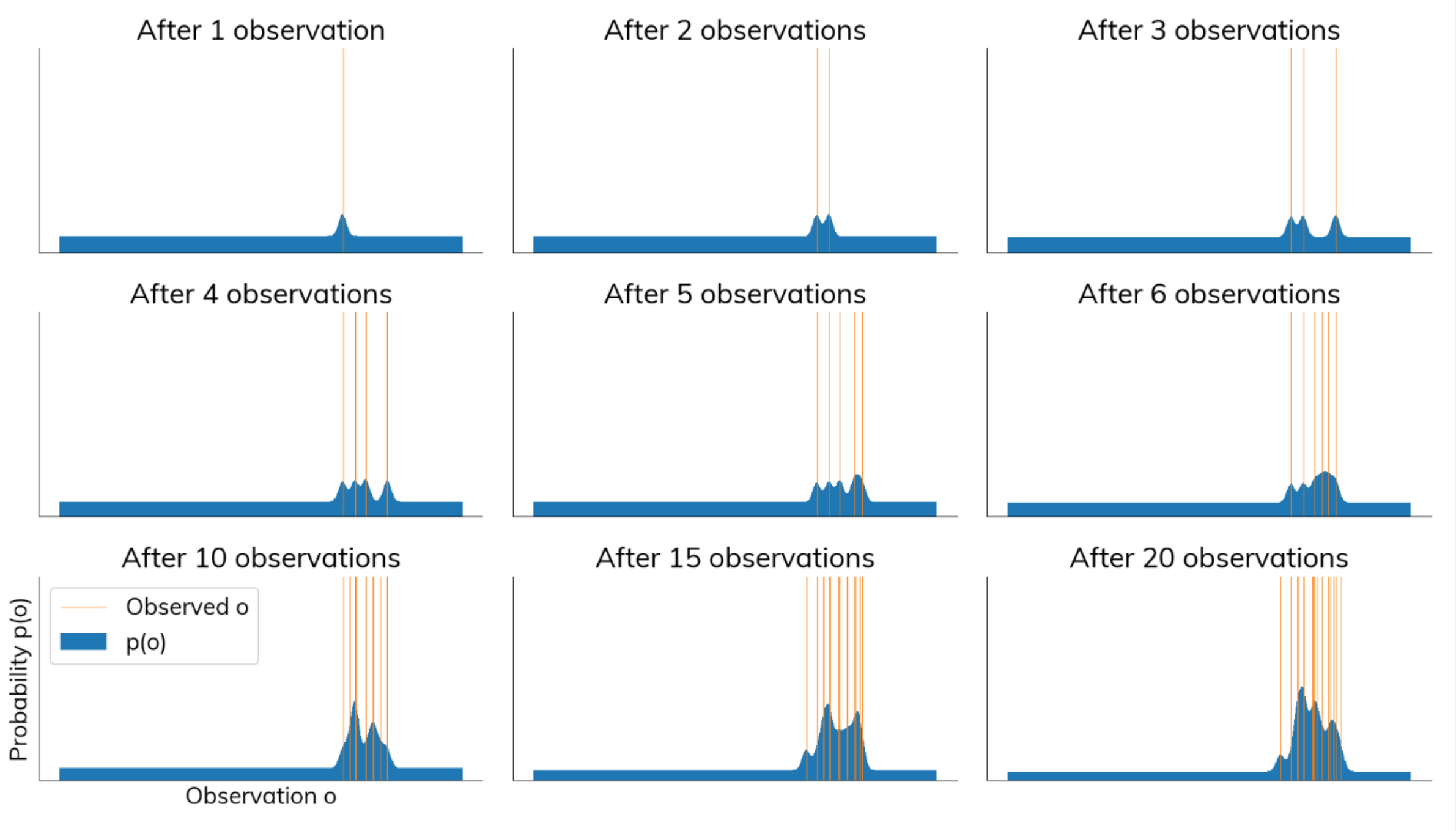

**Figure 4.** A slightly more sophisticated approach: approximating the probabilistic structure of observations by having each observation introduce some probability mass that spreads in its vicinity.

More canonical "Bayesian brain" narratives might focus here on the fact that the data are apparently drawn from an underlying (normal) distribution with some mean and standard deviation, and that the brain's task is to figure out the posterior distribution over the parameters of this distribution $p(\mu,\sigma|o)$. In other words, it might be said, the brain is trying to figure out the latent parameters $\mu$ and $\sigma$, while presented only with the observations o. These latent parameters then reflect an external 'hidden' state which is responsible for causing the observations, and the brain is tasked with Bayesian inference to figure out the most likely such cause. This type of inference to hidden states would indeed achieve an efficient coding scheme (since probability mass is allocated to actually likely observations), but it is important to emphasize that positing hidden states here would be (an interpretation of) a consequence of the general principle of efficient coding of incoming observations. We, as external observers, may describe the system as inferring hidden states, in this special case, but there is no need for that: the mere existence of structure in the observations is sufficient for there to exist some more efficient coding scheme than that corresponding to the uniform distribution, which systems can track and exploit. In this example I am explicitly emphasizing that each observation has some *nonparametric* influence over the agent's current approximation of p(o). In this sense, my proposal is akin in spirit to the "direct fit to nature" perspective on biological and artificial neural networks that was recently put forward (Hasson et al., 2020): neural networks "do not learn simple, human-interpretable rules or representations of the world; rather, they use local computations to interpolate over task-relevant manifolds".

The probability distribution in Figure 4 was built up by each observation imparting an equal amount of probability mass, i.e. each observation received equal weight. (Such an approximation scheme is typically referred to as a kernel density estimate (Rosenblatt, 1956).) This is appropriate and will yield a good approximation of p(o) when observations are made with equal, high reliability, and when the space of possible observations is limited and known (in this case, only 1,024 observations are possible). In reality, however, neither of these simplifications hold. First, observations are typically noisy, or, in any case, of varying reliability. Second, the space of possible observations is high-dimensional and extremely large. No real organism can ever hope to construct a true probability distribution p(o) over the entirety of such a space. This means that, if an organism were to use something like the above scheme, any new observation $o_t$ will, overwhelmingly likely, fall in the part of p(o) that has not been updated with any bits of probability mass in the past. In other words, it is very unlikely that any new observation (across all sensory inputs) is equal to any past observation, or even close enough for the above scheme to result in any appreciable increase in efficiency (or decrease in required energy). Therefore, our agent cannot exactly use the simple "kernel density" approach of approximating p(o) sketched above; the past needs to have a more sophisticated influence over the future. How can our agent best focus its lens through which observations are encoded, in order to capture as much meaningful information as possible, while accounting for the noisy and high-dimensional nature of its input?

## 2.4 Priors and regularization are two sides of the same coin

The field of machine learning is very familiar with high-dimensional spaces and noisy observations. The quintessential technique for dealing with the issues arising from them in this field is *regularization*. Phrased generally, if we want to fit a model f(x) to some collection of data points x, then regularization techniques force a solution that captures as much meaningful variation in the data as possible, but as little noise. In other words, regularization techniques govern the tradeoff between how much variance in the data to fit, and how much bias to introduce in the provided solution. For an intuition, see the following four regression models fit to the exact same twelve data points (Figure 5):

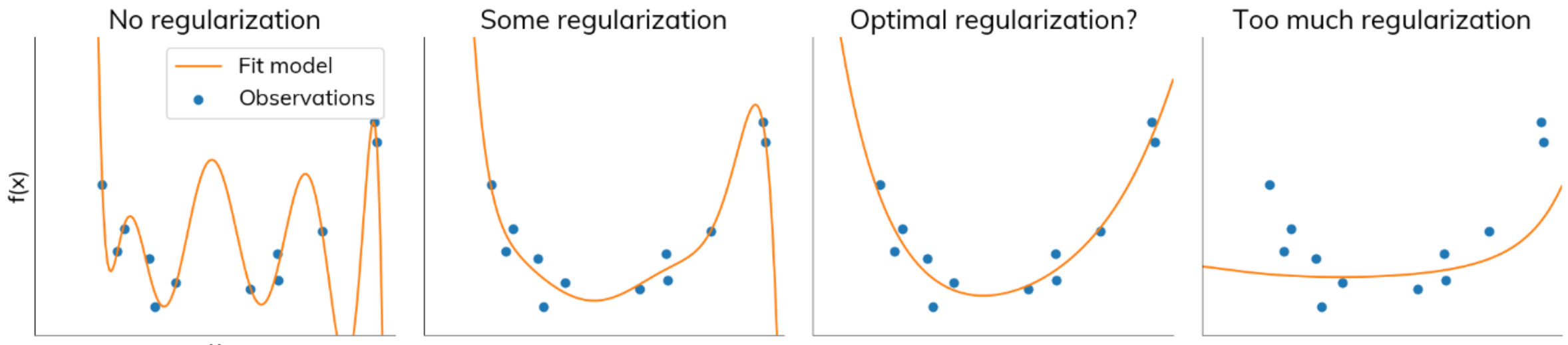

**Figure 5.** Four models fit to the same twelve data points. The leftmost panel shows a perfect fit, but does not capture any meaningful pattern. The data have pulled the model too strongly towards them; more bias is needed to counteract the noisy data. In contrast, in the rightmost panel, the data appear to have almost no influence over the model. The optimum is somewhere in between.

The leftmost regression model in Figure 5 fits the data perfectly! However, it very likely does not capture any meaningful pattern that would translate well to new future observations. We can say that the data have exerted too strong an influence over the model. This influence needs to be counteracted by a regularizing bias. In the rightmost panel, we can see the effect if this bias is too strong: the data are hardly reflected in the model anymore. The optimal regularization strength is not knowable *a priori* without knowing the underlying true model or signal-to-noise ratio of the observations, but it likely lies somewhere in between the two extremes.

Regularization in optimization problems is typically implemented by adding a penalty term to the function to be optimized, such that too complex solutions (such as the leftmost panel in Figure 5) are discarded because the corresponding penalty term is then very high. For the present narrative, another view of regularization is important: it turns out that regularization is equivalent to Bayesian inference, where the priors over parameters determine the type and strength of regularization (Berger, 1985; Figueiredo, 2003). The form of regularization used in Figure 5 (very commonly used and known by many names including "shrinkage" regularization and ridge regression) includes a penalty term that forces model coefficients to be small. This type of regularization is specifically equivalent to a Gaussian prior over the regression weights. Such a regularizing prior has the effect of *pulling* model coefficients towards itself, thereby bringing the optimization solution more in line with what we expect a good solution to look like. Graphically, we can depict this pulling force as done in Figure 6:

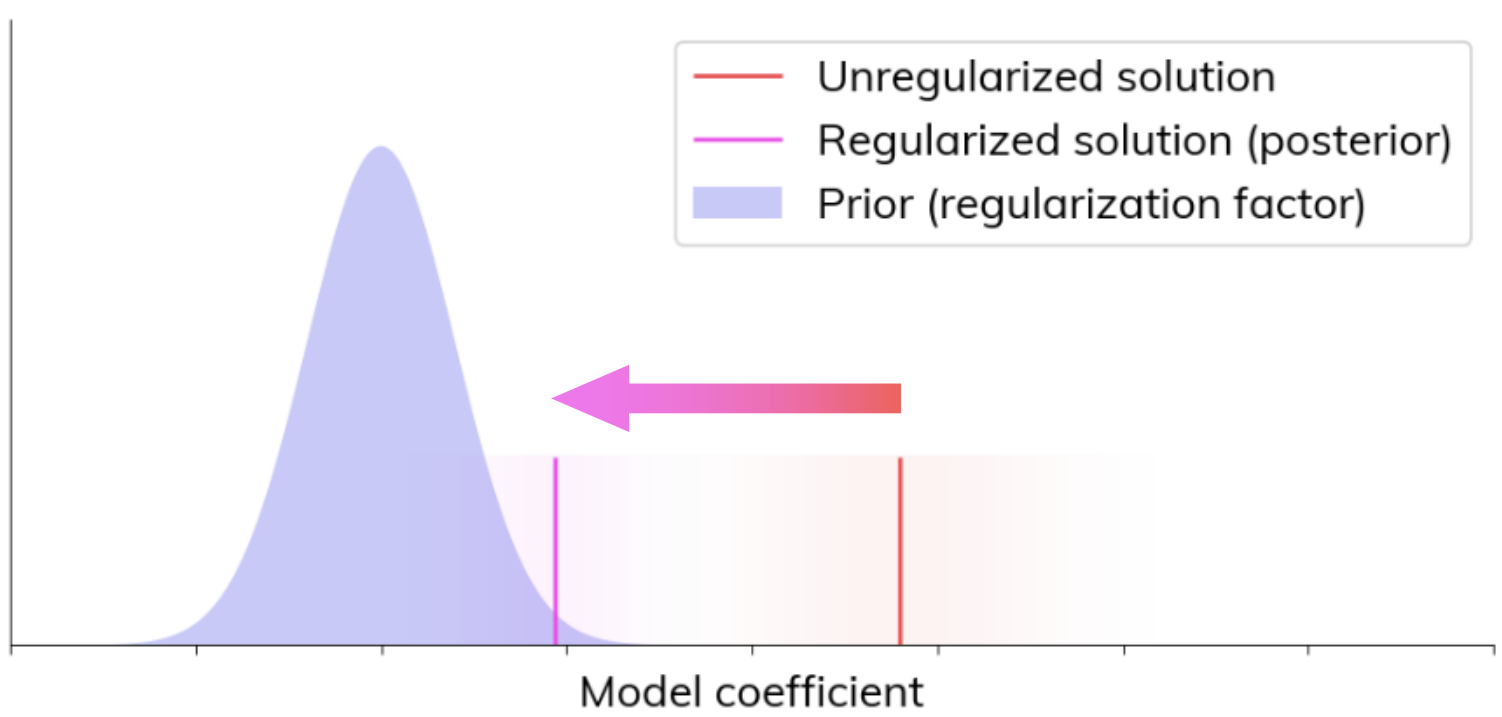

**Figure 6.** The difference between an unregularized model (red) and a regularized one (magenta) is the introduction of a prior factor (a bias; lavender) that pulls the coefficients towards it.

Just as regularization factors in machine learning can be interpreted as priors in Bayesian inference, so can we interpret prior expectations over sensory observations as regularizing factors for encoding newly incoming sensations. Any stimulation of sensory cells exerts a 'tug' on the system that is our agent, just like data points tug on models in machine learning. If data points are allowed free reign to tug as much as they like, we end up with massively overfit solutions like the one in the leftmost panel of Figure 5. Instead, a counteracting force is needed in order to extract as much *useful* information from the data as possible: the regularization. Sensory observations, with their noisy nature and massive dimensionality, similarly should not be granted unlimited tugging strength: the past, encoded in the prior p(o), should exert a pulling force as well.

It is worth emphasizing here that there is *not* first some agent-independent observation $o$ which is subsequently encoded by the agent into something like $o_{\text{encoded}}$, and that the prior only then comes into play, tugging on the observation. Generally, an observation is some influence that the world has exerted on the agent. The very nature of this influence depends not just on the world, but also on the agent; specifically, the prior distribution p(o) and the encoding scheme that follows from it determine the consequences for the agent of various perturbations imparted on it by the world.

Again, we can view the regularizing consequences that our agent's prior p(o) has for an incoming new observation $o_{21}$ graphically, as in Figure 7:

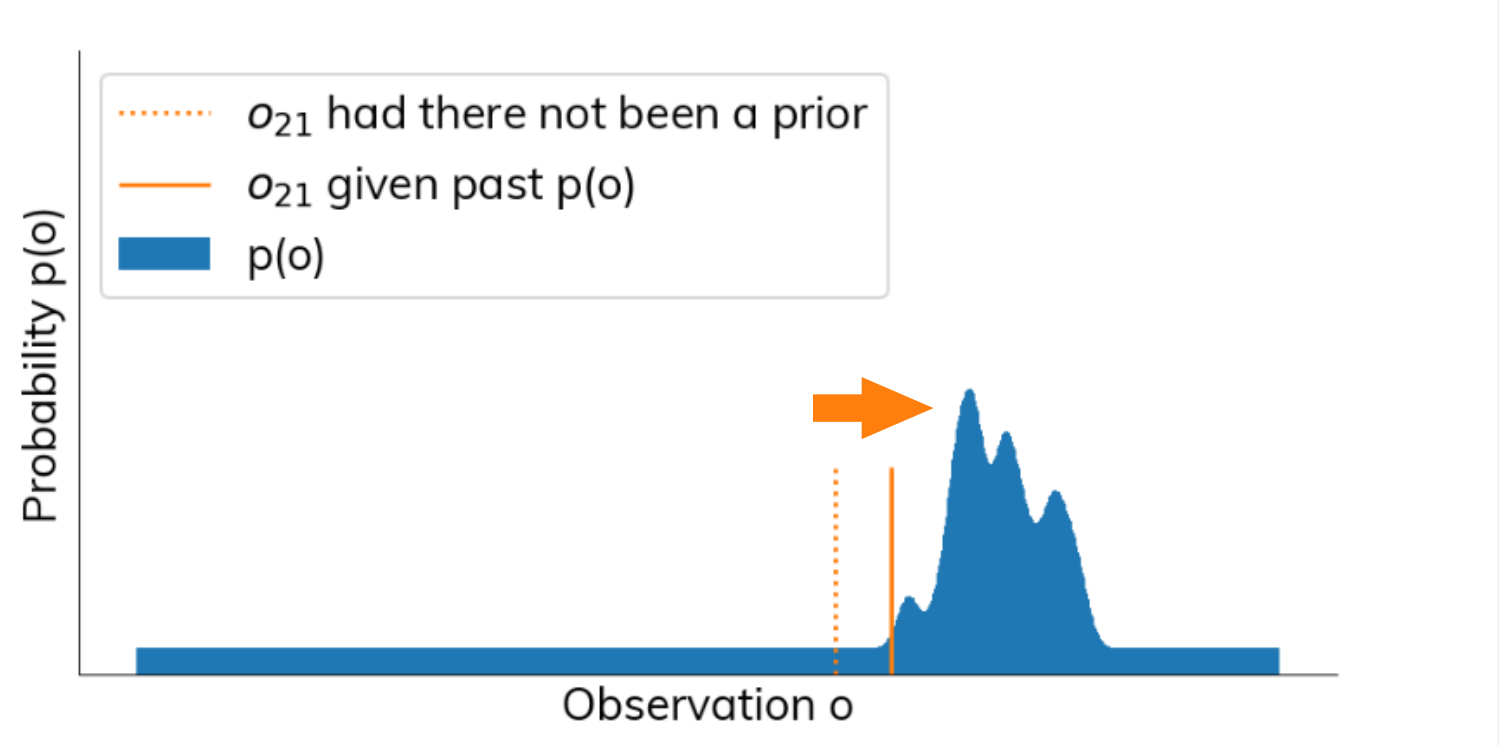

**Figure 7.** Priors of arbitrary complexity, such as those corresponding to the expected statistical structure of observations, also exert regularizing pull.

(In turn, of course, this new encoded observation $o_{21}$ will slightly change p(o) for when a later $o_{22}$ comes in, et cetera.)

## 2.5 Summary so far

In order for an organism to extract as much useful information from the noisy and vastly multidimensional incoming stream of observations as possible, it is necessary for it to encode this information parsimoniously. The apparent stability of the world suggests a natural option for such a parsimonious scheme: leverage and track the probabilistic structure of observations p(o). High-probability regions of an agent's approximation of p(o) exert a regularizing pull on the encoding of new observations, which in turn serve to structure the tracked p(o).

The notion of 'pull' or 'tug' here thus works both ways. The following example illustrates this nicely. Humans and other primates typically encounter visual sensations stemming from faces very often. This exposure leads to an accumulation of probability mass in the 'face' region of p(o); frequent similarly-structured stimuli pull probability mass towards them (although debated, note that face preference indeed appears largely learned and not innate (Arcaro et al., 2017)). In turn, the resulting bump in p(o) pulls subsequent somewhat similar observations towards it, to sometimes comic effect (Figure 8):

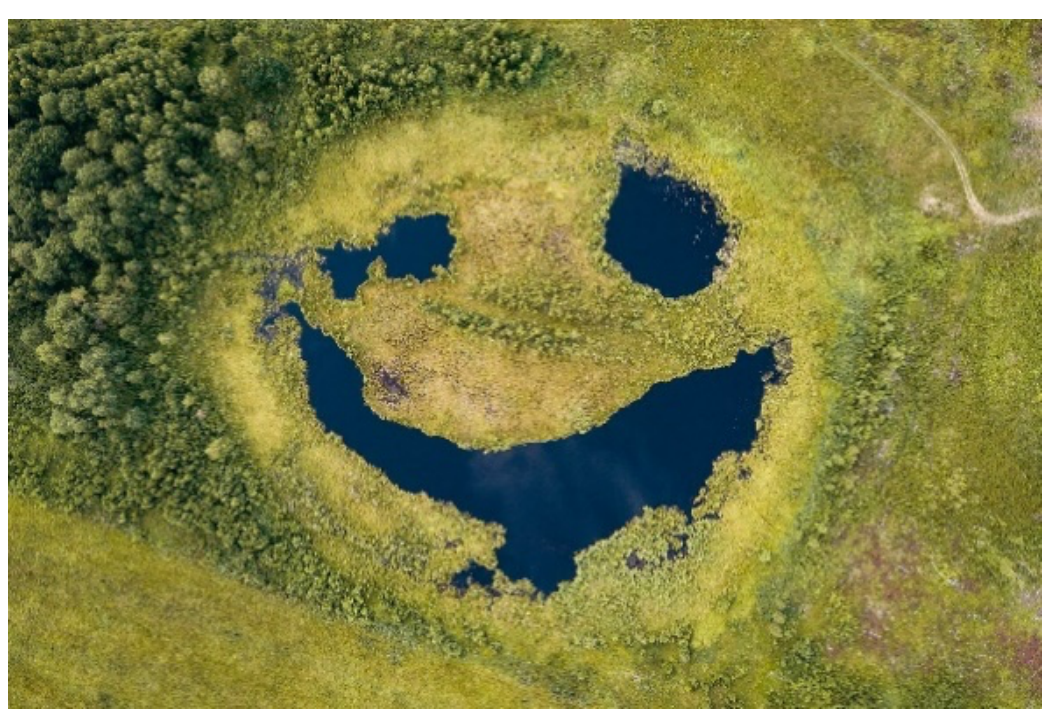

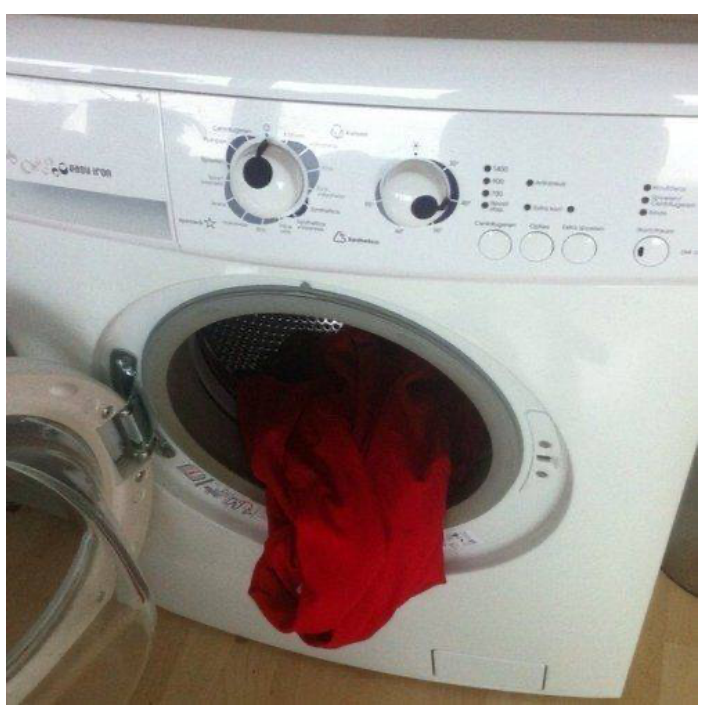

**Figure 8.** A group of meteor lakes in Russia (left) and a washing machine (right).

## 3. Action generation through minimizing model mismatch

### 3.1 From perception to action

In most theories adopting the Bayesian brain framework, focus lies on the perceptual tasks that organisms are faced with. Also in the present work, so far, our agent's task has been phrased in terms of "encoding observations", which has a strong perceptual connotation. We now reach a point where some interesting (conceptual) work can be done by us deliberately avoiding the "inferring hidden causes of observations" angle on neural Bayesian inference. Instead, I have described the bidirectional pull between the previously established best approximation of p(o) on the one hand, and influence from the world, encoded largely through the lens of p(o), on the other. Throughout this process, the agent is continuously optimizing the way it encodes p(o), so as to best accommodate new observations *o*. In essence, it strives to minimize the mismatch between its model of p(o) and newly incoming observations *o*.

We are interested in *neural* systems, so this optimization process has to happen through changes in firing patterns and connectivity states. Talking about probability distributions is metaphorical for what the brain is computing; physically, all there is is tissue. Updates in certain parts of the nervous system, such as visual sensory cortex, might be very straightforward to interpret for outside observers as the organism discovering ever higher order categories that can be extracted from the stream of visual input. Retinal light patterns can be parsed into edges, which can be parsed into textures, which can be parsed into objects. This perspective on visual cortex as a feedforward feature extraction machine has been instrumental in our understanding of brain function in general. Importantly, this perspective also lends itself well to a 'Bayesian inversion': objects induce expectations over textures, which induce expectations over edges, which induce expectations over retinal light patterns. The ease by which we can understand this inversion is, I believe, part of the reason why Bayesian theories of brain functioning have been so influential.

However, we should not forget that such an apparent hierarchy of inference is only one way by which the brain can minimize the mismatch between its approximation of p(o) and incoming *o*. The overall goal is the minimization itself. Changes in neural firing or connectivity anywhere in the nervous system might contribute to this objective. If these changes occur in parts of the nervous system that happen to be connected to muscle tissue, then overt behaviour ensues.

Imagine that our agent wants to sit down, but is currently standing. Because it is standing, the proprioceptive observations that are arriving at the nervous system are compatible with standing. If the above machinery of minimizing mismatch between the landscape of p(o) and incoming observations is in place, we can conceive of any desire as a region of high 'probability'[4] mass in p(o). In this case, the desire for sitting would correspond to a bump of high 'probability' specifically for observing the proprioceptive consequences of sitting. This bump will exert its pull on the incoming observation *o*, just as we described for perception. In order to minimize the mismatch between *o* and p(o), again updates in neural firing and connectivity will happen; in this case, the 'best' updates to execute will have consequences for muscle activity, which, in turn, influences skeletal posture, which, in turn, influences proprioceptive observations, thereby achieving the minimization of mismatch and fulfilment of desire.

[4] I am writing 'probability' here in scare quotes because this is stretching the metaphor of probabilistic inference to (and arguably beyond) its limits. Please bear with me; I return to this shortly.

There are three important takeaways from this perspective on action generation.

### 3.2 Behaviour is primarily the result of controlling *input*

First, action is a consequence of approximating the structure of *observations* by an internal model p(o). The actual generation of behaviour, output being sent to muscles or other actuators, is corollary to the system minimizing the mismatch between p(o) and *o*. Whereas in the sensory narrative, it is natural to emphasize the changes in p(o) in this optimization scheme, in the motor narrative, changes in the (encoded) *o* are more relevant. Systems equipped with actuators can minimize mismatch between input and internal state not only by adapting the internal state, but also by firing actuators. At least in biological brains, there is no principled difference: internal states as well as muscle outputs are both counterparts of patterns of activity in neurons. The ultimate goal of the brain is to control the body, but this control is achieved by characterizing desirable *input* states. This idea has a predecessor in the work of Powers, who similarly viewed all behaviour as the control of perception: "control systems control what they sense, not what they do." (Powers, 1973; Seth & Tsakiris, 2018).

### 3.3 Approximated structure of observations is highly dynamic

Second, we see now that the pulling force of islands of probability mass in the tracked p(o) has to be dynamic and variable. There is a clear and distinct difference between wanting to sit down, and believing that you are sitting. It does not make sense to say that a standing agent already believes itself to be sitting before it has engaged the behaviour that would make it so (see also Yon et al., 2020). While beliefs and desires are different cognitive constructs, we can nonetheless appreciate that they share an underlying mechanism. Both correspond to regularizing factors, to bumps in an agent's approximated p(o). Some of these bumps are best interpreted as desires (presumably, steep bumps that have a relatively proximate link to generating behaviour), while others are best interpreted as beliefs (shallower bumps for which the link to generating behaviour is much more indirect). Oversimplifying the distinction between the two, we can view these two different pulling effects as in Figure 9:

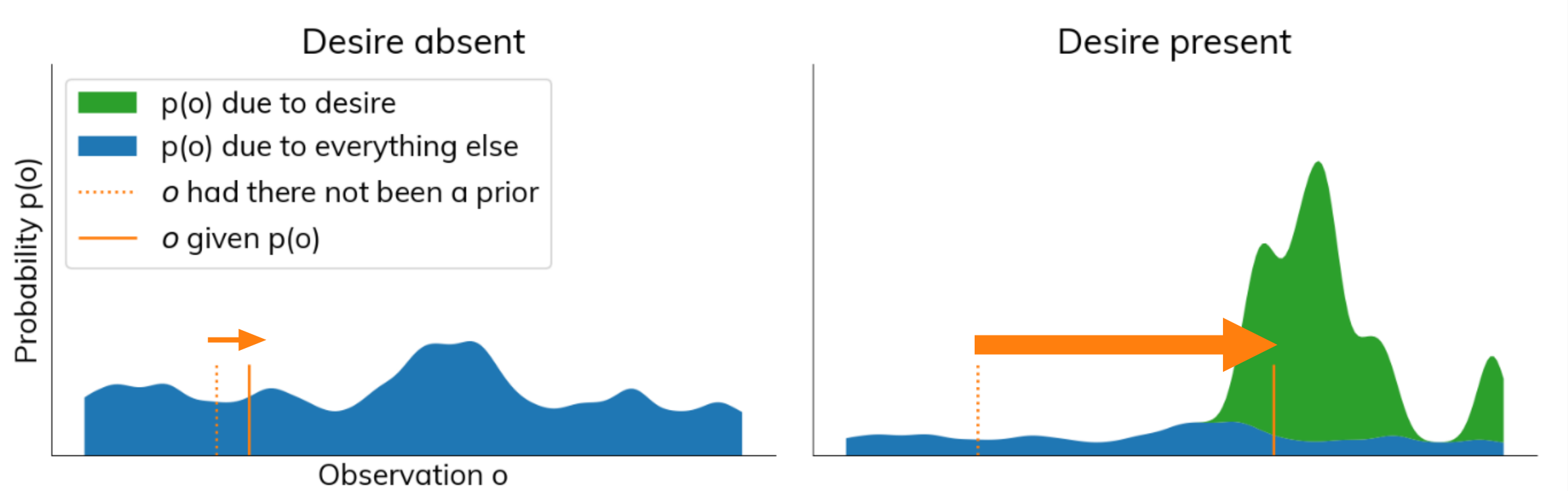

**Figure 9.** Different factors that go into an agent's approximated p(o) can have different strengths. On the left: the majority of factors exert a moderate pull on encoding new observations. On the right: one factor in particular, corresponding to a desire, can be thought of as exerting a strong pull, which is specifically not satisfiable through any model updates but instead requires actuators firing.

At any time, the agent's approximation of p(o) is built up from many different factors. In Figure 9, I have lumped all those factors, except for one, together in the blue probability curve. The resulting p(o) from all these various factors might result in a moderate pull on the encoding of any incoming observation, as depicted on the left. However, when a desire is present, the landscape of p(o) now suddenly has a steep hill (arguably of a qualitatively different nature): this amounts to a significant amount of 'probability' mass that corresponds to the desire to sit (or more properly: the desire to perceive the consequences of sitting). The pulling force that this hill now exerts can only be resolved by actually changing the inputs to proprioceptive sensors, which in turn is achieved by muscle movement.

### 3.4 'Probability matching' is a metaphor and therefore has limited applicability

As a third and final takeaway from this perspective on action generation: readers might appreciate that we are running into the limits of the probability distribution metaphor here. Probability distributions over some quantity are meant to capture beliefs and expectations about that quantity; using them to describe desires for action requires significant shoehorning. It is possible to bite the bullet and say that agents "expect their goals to be fulfilled", an approach which is often taken by proponents of so-called active inference and the free-energy principle (Friston et al., 2015). While I am sympathetic to these proposals and their unifying goals, I am reluctant to bite that bullet and have to deal with the flagrant betrayal of common sense that follows. The computational (physical) machinery that draws an agent towards a particular state is likely similar whether that state corresponds to an optimal approximation of statistical structure, or whether it corresponds to a to-be-achieved goal state. However, literally re-interpreting desires as nothing other than beliefs is confusing and might, I believe, harm the acceptance of this type of unifying theory of brain function.

A particularly salient example of the above phenomenon is the free-energy principle account of homeo- and allostasis. Uncontroversially, organisms need to keep certain parameters, like blood carbon dioxide levels, within tight bounds; bounds which are governed by homeo- and allostatic machinery. One might posit that organisms "*a priori* expect their physical states to possess key invariance properties […] mandated by the very existence of agents and [leading] naturally to phenomena like homeostasis" (Friston, 2011). It is indeed true that, if I think about it, I expect my blood carbon dioxide levels to be within bounds compatible with life, but I have this expectation because I (consciously) know that these levels *need* to be bounded. Homeostasis is a consequence of a biological *need*, not an expectation. Through the lens of regularization: the need for $CO_2$ levels to be within certain bounds is best captured by a regularizing factor so strong that the 'solution state' is (almost) entirely determined by the regularizing bias, and the data have (almost) no influence (as e.g. in the rightmost panel of Figure 5). An even better term here would be a *constraint*, which is a special case of a regularizing factor. As long as the constraints of bounded $CO_2$ levels are not satisfied, these will exert pulls on the state of the organism which will always result in behaviour, since no internal model updates are possible which would satisfy the constraint instead.

To further illustrate the harm that sticking to the language of beliefs and expectations well beyond perceptual inference can do, consider the 'dark room paradox' (Sun & Firestone, 2020). At its core, the paradox is simple: if the overarching goal of the brain is to minimize the mismatch between expected and actual sensory states, then why do organisms not all crawl into dark, silent caves, while expecting to hear and see nothing? After all, that would categorically minimize mismatch and thus be a means of achieving this goal. This 'solution' is clearly absurd, so the premise must go. I contend that the problematic part of the premise is in the labelling of priors as 'expected'. It is trivial for us to explicitly expect a certain state of affairs, and if that state of affairs comes about, then our expectations are fulfilled. This common-sense use of the word 'expect' carries with it the seeds that germinate into the dark room paradox. When it comes to an agent's internal model of the (probability distribution of) incoming sensations, there is no way to explicitly reorder it to 'expect' any arbitrary incoming sensation. Instead, this model is always structured and comprises, among other things, (hard) constraints such as those corresponding to homeostasis. Therefore, crawling into dark, silent caverns and allocating 'probability' mass accordingly can never be a solution to the overarching goal of minimizing mismatch. This state violates hard constraints (think of hunger, thirst), thus strong pull remains exercised by peaks elsewhere in the 'probability' landscape.[5]

### 3.5 Organisms are never *tabulae rasae*

It is perhaps self-evident and already hinted at in the previous paragraph, but worth emphasizing again that no organism is born as the kind of *tabula rasa* agent which I sketched at the beginning of section 2. The structure of the approximated p(o) implicit in a newborn organism's state is probably a lot less lumpy than that of an adult, but still far from uniform. Instead, many factors already govern the way new *o* are encoded (equivalently: govern the system's response to incoming perturbations), including critical ones like those corresponding to homeostasis. Organisms without such 'priors' in place would not survive and therefore could not have evolved. We may consider 'priors' less directly essential for life than homeostasis as constituting some 'reward function', and this, too, is likely (indirectly) rooted in the optimization of evolutionary fitness (Singh et al., 2009).[6] In that sense, the structure of the world exerts a causal influence over approximated p(o) that is not limited to an individual agent's past observations; instead, it stretches billions of years into the past.

[5] The idea that the brain is continuously trying to reason from sensory effects to (hidden) worldly causes has been compared to the way in which scientists reason from observed data to (unobserved) theoretical constructs. Recently, this picture of an 'ideal' scientist has rightly been criticized by noting that the brain likely embodies several 'beliefs' that are never amenable to updating in light of new, conflicting evidence (such as the belief that 'I am alive') (Bruineberg et al., 2018; Yon et al., 2018). In spirit, this dovetails well with the view I am espousing here, though I emphasize how some of the problematic work is done by sticking to the language of 'prior beliefs' in the first place.

### 3.6 Summary of action generation, introducing attractors

Let us summarize the account of action generation from this perspective on the Bayesian brain. Once the machinery of minimizing mismatch between an internal approximation of p(o) and the actual structure of incoming *o* is in place, and we appreciate that many different factors contribute to the form of the approximated p(o), we can see that desires and even "hard-wired" constraints can function analogously to prior expectations. They exert a pulling force that determines the state of the organism, in concert with ongoing perturbations form the world. Such forces always act on incoming sensations, and there is no principled difference between optimizing the approximation of p(o) and the resultant encoding of observations through neural changes interpretable as model updates on the one hand, and achieving that same goal through neural changes better interpretable as behaviour on the other. Behaviour is but one way – an often very efficient one – by which the organism can control its (encoded) inputs.

Priors, (efficient) coding schemes, regularizers, constraints: these turn out to be different perspectives on the same underlying mechanism. The notion of prior expectations is well aligned with our common sense and scientific thinking about perceptual systems, whereas constraint satisfaction offers a more natural analogue when describing the generation of behaviour. Due to the shared underlying machinery of mismatch minimization, we can freely toggle between these perspectives (to varying degrees of intuitional controversy). We can describe the regularizing effect of behavioural goals as prior expectations, as is done in work on active inference (Friston et al., 2015). Conversely, we can interpret prior sensory expectations as regularizing factors for incoming data, as I have done above. Throughout the present narrative, the notion of an abstract 'pulling' force has played a central role. This suggests a natural interpretation of all these three concepts: the *attractor*.

[6] The general question of 'where goals come from' in this narrative is an interesting one. While I speculate here on their ultimate origin in evolutionary fitness, a thorough treatment is beyond the scope of this paper. Here we are mainly concerned with understanding how goals can be integrated in the general framework of probabilistic inference over observations.

## 4. Attractors unify priors and regularization, goals and constraints

In this section, I will first briefly introduce the concept of attractors in dynamical systems in general, as this is likely the concept in this narrative that cognitive (neuro)scientists will be least familiar with. (Although note that fruitful prior work in cognitive science exists that draws explicitly on dynamical systems theory, e.g. (Beer, 2000; Smith & Thelen, 2003).) After that, I will outline how this physical concept might unify the various regularizing forces underlying perceptual inference and action generation that make up an organism's best approximation of the statistical structure of its observations.

### 4.1 Attractors in dynamical systems

Attractors are an important concept in the study of dynamical systems, i.e. those systems that change over time. Dynamical systems are typically modelled either by differential equations for continuous time (i.e. $dx/dt = f(x)$) or iterated functions for discrete time (i.e. $x_{t+1} = f(x_t)$). A one-dimensional dynamical system is described by a point which moves about on a line, and can be characterized by noting how this movement changes for different regions of that line. Figure 10A shows this rate-of-change curve for a simple 1D system:

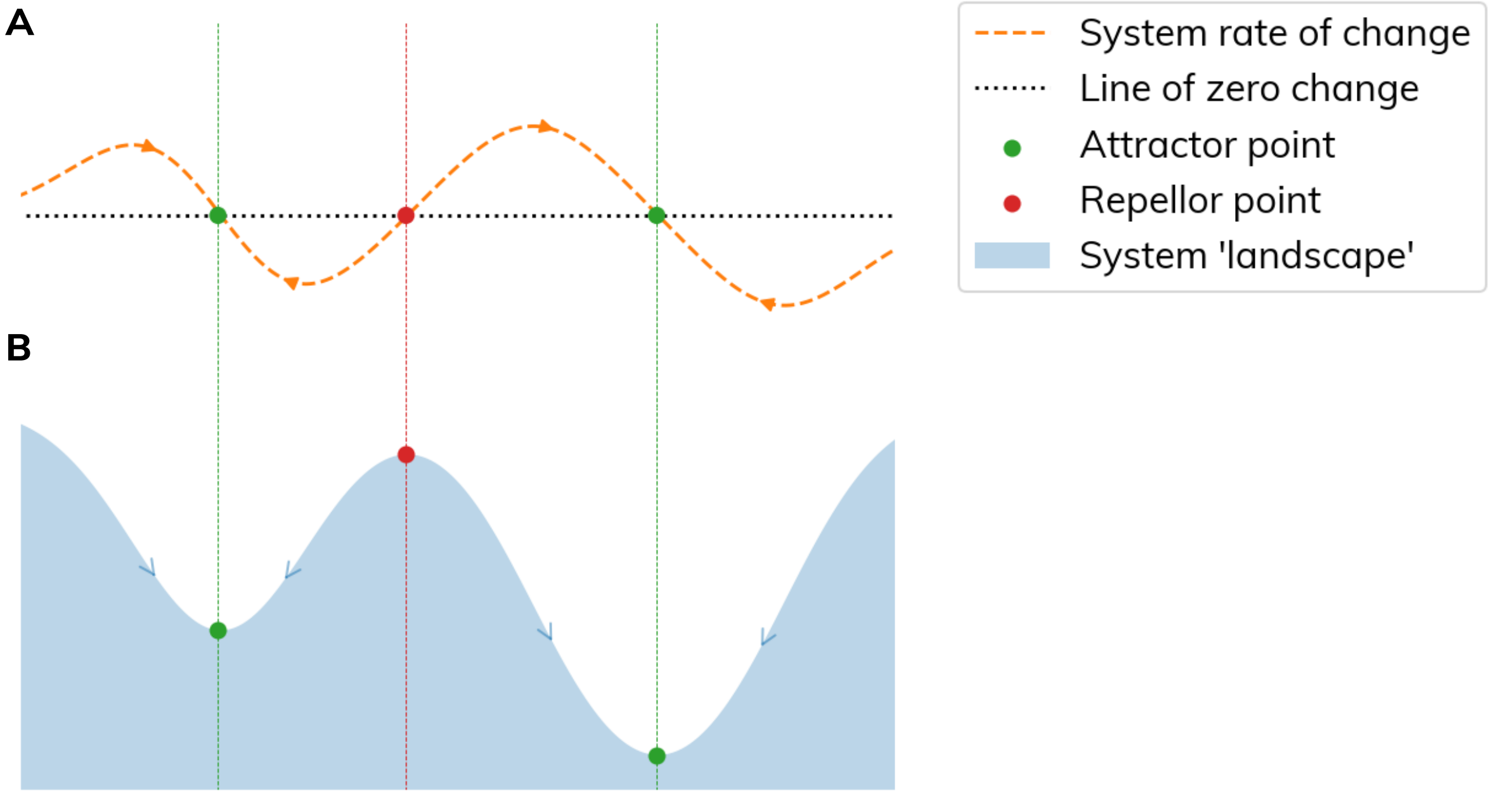

**Figure 10.** A simple one-dimensional dynamical system. **(A)** The curve that governs the rate of change of this system. **(B)** Visualizing the same system as a landscape in which a ball rolls around.

Whenever the system reaches a zero crossing in the rate-of-change curve, it will remain stationary from then onwards (which is the definition of zero change). Such zero crossings are called fixed points of the system. Fixed points can be repelling or attracting. If the system starts in the vicinity of a repellor (the red dot in Figure 10A/B), it will proceed to move further away from that repellor. In contrast, if the system enters the vicinity of an attractor, it will continue to evolve in the direction of that attractor. Attractors are thus possible states of a dynamical system to which other states of the system tend to evolve.

We can visualize the dynamical system in Figure 10A as a hilly landscape, shown in Figure 10B. If we were to perch a ball exactly on top of the hill indicated by the red dot, it would stay there, since the rate of change curve tells us that this point corresponds to zero change (i.e., the hilltop is a fixed point). However, if we were to put the ball even the tiniest bit to the left or right of the red dot, it would proceed to roll into either the left or right valley (i.e., the hilltop is a *repelling* fixed point). It would eventually settle at the lowest point indicated by the two green dots, which correspond to *attracting* fixed points. A ball dropped anywhere in the left or right valley would always settle at the same left or right attractor; the exact initial condition (other than which valley it corresponds to) does not matter for the final state of the system.

The only type of attractor possible in a one-dimensional continuous-time system is the attracting fixed point; the only stable solution in 1D is for the system to stop evolving altogether. In two dimensions, more interesting attractors are possible, such as periodic cycles (also known as 'limit cycles'). 2D dynamical systems are no longer describable by considering a point on a line, but instead by a point that moves about in a plane. Figure 11 shows a 2D dynamical system that has a periodic attractor (known as the 'Van der Pol' oscillator):

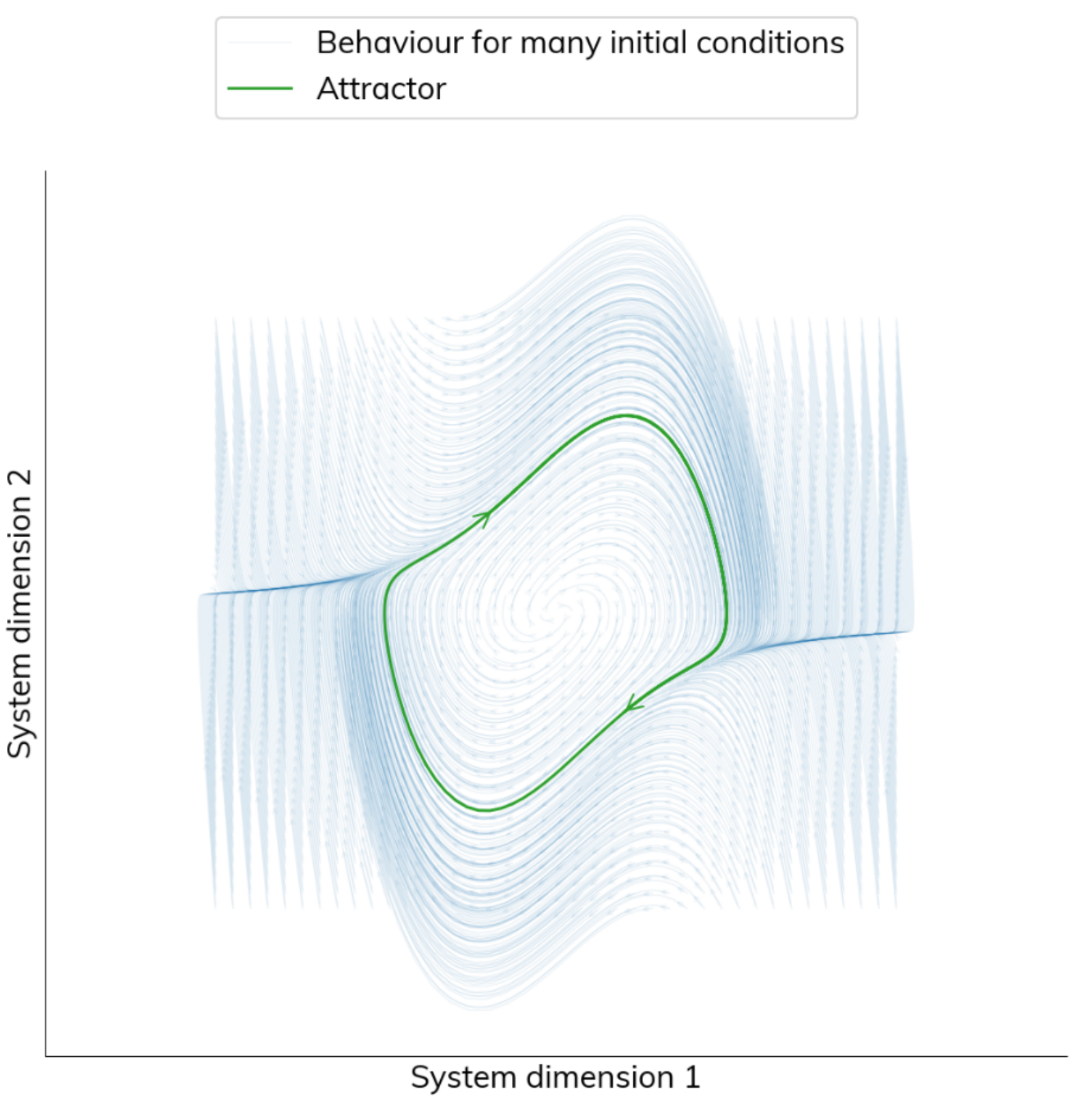

**Figure 11.** A two-dimensional dynamical system that has a periodic attractor.

In Figure 11, the thin blue lines correspond to evolutions of this system for many different possible starting points. All states of this particular 2D system eventually get pulled towards the cycle depicted in green. This curve therefore is an attractor of the system, even though it is not a fixed point. Wherever the system happens to start out, it tends to evolve towards cyclic behaviour on the attractor, and once it's there, it stays there, forever tracing the same cycle.

### 4.2 Attractors can be chaotic

In two dimensions, the only types of attractors that are possible are fixed points (like we saw for one dimension) and periodic limit cycles. In the back of your mind, you might be wondering: if the idea is to apply the attractor concept to states of an organism and its brain, then these attractor types are highly simplistic. Clearly the brain does not evolve towards a stationary fixed point, and neither does it keep evolving in the exact same circular orbit. I will have more to say about this soon, but before I do it is important to introduce a third type of attractor, which might already alleviate some of this worry from simplicity. As we increase the dimensionality of our system to three or higher, we now also see the appearance of *aperiodic* attractors. Like all attractors, an aperiodic attractor is a region in the space describing the system's behaviour to which all other states tend to evolve. Unlike fixed points or limit cycles, however, once a system reaches the orbit of an aperiodic attractor, the behaviour while on that attractor keeps evolving along unique trajectories. The exact trajectory which the system follows on such an attractor is sensitively dependent on the initial state of the system: even minute differences in starting points result in very different trajectories of the system. Nonetheless, all those trajectories are pulled towards the attractor, and their unique paths are traced on the attractor, and not outside of it. Because of its sensitive dependence on initial conditions, this type of attractor is sometimes also called a *chaotic* attractor. The upshot here is this: even though dynamical systems can have a clear and strong attractor structure, in higher dimensions ($\geq 3$) this can still lead to arbitrarily complex (and not trivially predictable) behaviour. Figure 12 shows a three-dimensional system (now describable by tracing a point in 3D space) that houses an aperiodic attractor (the 'Lorenz' attractor):

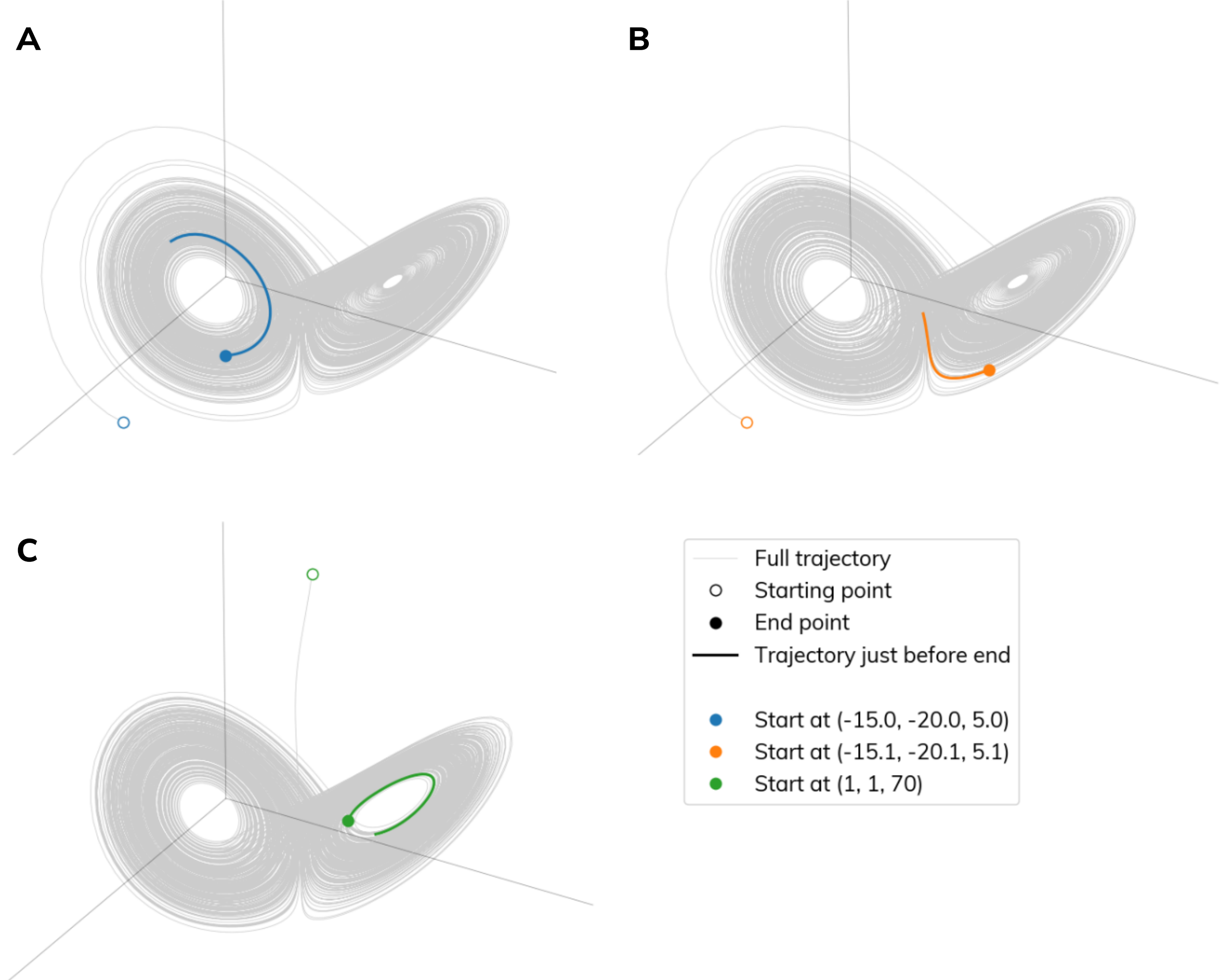

**Figure 12.** A three-dimensional dynamical system that has an aperiodic attractor. Panels A-C show different starting conditions of the same system (coloured open circles). For all possible starting conditions, including widely varying ones such as A and C, the system traces out the same attractor shape (light grey). At the same time, even for starting conditions very close together (A and B), after some time elapses, the trajectories become unique and are not predictable from the starting point.

## 4.3 Approximated p(o) as an attractor

We have seen that agents can approximate and use the probabilistic structure of incoming observations in order to efficiently encode those observations and extract useful information from them. This approximation takes the form of regions of higher past probability receiving additional bumps of probability mass for the future. These bumps then exert a pulling force on newly incoming observations (which in turn result in new bumps, et cetera). We can view this pull as a prior in Bayesian inference, yet we can equivalently view it as a regularizing factor, or, if the pull is strong enough, as a constraint to be satisfied. The bi-directional pull between these factors and (encoded) incoming sensations has consequences in the form of internal model updates and/or behaviour.

It is well established that the behaviour of individual neurons can be fruitfully described and analyzed by viewing them as dynamical systems (Izhikevich, 2010). A full nervous system, or even an entire organism, is difficult to analyze using mathematical or numeric techniques, due to its highly nonlinear and multidimensional nature. Nevertheless, we can still appreciate that they are dynamical systems, systems that change over time. Therefore, general principles applying to all dynamical systems also apply to these biological systems of interest. In particular, the space of states for our nervous systems might house (chaotic) attractors. In fact, it turns out that even relatively small and simple neural networks with balanced excitatory and inhibitory connections can display chaotic structure and complex attractors (Miller, 2016; Sompolinsky et al., 1988; van Vreeswijk & Sompolinsky, 1996; Yang & Huang, 2006), so the existence of these in a full biological nervous system is not just possible, but likely.

We can easily define an input space of a biological (nervous) system, for example by assigning one dimension to each sensory cell. Per the foregoing, the brain's overarching goal is then to track the probabilistic structure of this space of inputs in order to efficiently encode sensations and generate action. Efficiently encoding sensations is achieved by allocating Bayesian priors that correspond to experienced history. These priors act as regularizing factors and thereby exercise an abstract pulling force on the incoming sensations. Action is likewise achieved by such a pulling force: factors that we could consider regularizing factors, desires, or, by stretching common sense, Bayesian priors, pull on incoming sensations. Attractors in a dynamical system exercise – by definition – a pulling force on the state of the system, and the existence of attractors in nervous systems is empirically and theoretically likely. Therefore, I propose that the 'bumpy' landscape I described previously, the set of Bayesian priors, regularizing factors, and constraints, exercising its pull on incoming sensations, is physically well-characterized as a (chaotic) attractor in the space of neural input states. In this way, the phenomena we defined from the perspective of information processing, probability, and teleology ("an organism *should* use efficient descriptions and minimize mismatch…") find a natural counterpart in physics.

### 4.4 Neural activity samples from a probability landscape

There is an interesting line of empirical work which demonstrates that neural activity in sensory cortex is well understood as reflecting a process of *sampling* from some probability distribution (Aitchison & Lengyel, 2016; Berkes et al., 2011; Buesing et al., 2011; Echeveste et al., 2020; Orbán et al., 2016). At any instant in time, the (high-dimensional) state of the neural system is one sample from an underlying distribution. Simplified, one might say that the activity of neuron 1 reflects the probability density along dimension 1 of some distribution, the activity of neuron 2 reflects the probability density along dimension 2, and so on. While the full distribution is never actualised in neural activity at any one instant in time; once aggregated over time, the path that the activity of the neural system traces does come to reflect the full distribution (Figure 13).

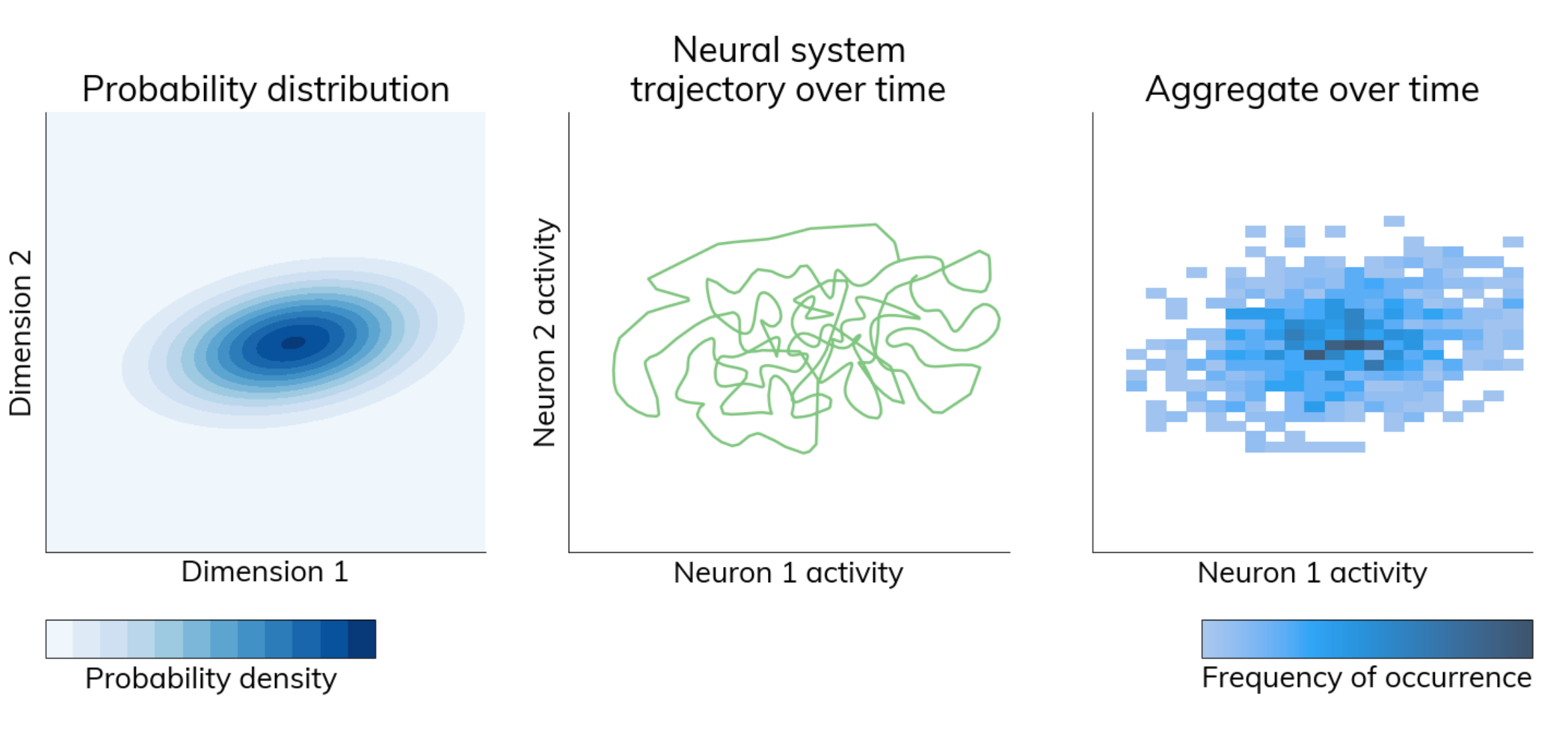

**Figure 13.** The neural sampling hypothesis. Neural activity in sensory cortex (middle panel) reflects sampling from an underlying probability distribution (left panel). As visible once aggregated over time, the neural system visits regions of its state space with a frequency proportional to the underlying probability (right panel).

The interpretation of sensory cortex neural activity reflecting samples from probability distributions dovetails nicely with the view of the landscape of approximated p(o) as an attractor in the space of states of the nervous system. When a system is producing samples from a probability distribution, at any random moment in time, we are most likely to find the system in regions of high probability. That is the definition of sampling: we should get more samples from high-probability regions than from low-probability regions. In that way, high-probability regions can be thought of as attractors in the dynamical system that is producing the samples. There is thus a clear analogy between a probability landscape that is being sampled from, and various priors/regularizing factors/constraints that are pulling system activity towards them.

There is an important but subtle nuance to be made in the analogy between typical attractors in dynamical systems and neural sampling from probability distributions. A physical attractor, by definition, is a region in the space of states of a system to which the system will evolve, for many different starting conditions, given enough time. Once on the attractor, the system will stay there, either stationary (fixed point), or while displaying periodic or chaotic behaviour. When sampling from a probability distribution, however, a system should not stay 'locked' in regions of high probability; instead, the entirety of probability space should be explored. In other words, if high probability regions are attractors, the system should be knocked away from these attractors as well as be drawn towards them. In the narrative of the aforementioned bumpy neural landscape as an attractor, I propose that this job of knocking away the system from attractors is fulfilled by the *world*: the world is continuously perturbing the system, never allowing it to settle[7]. While the attractor structure is apparent through the force it exerts on the

[7] *Prima facie*, the image of the world continuously perturbing the system may appear at odds with the inherent stability of the structure of the world across timescales, which I emphasized in section 2 as a prerequisite for probabilistic, purely observation-driven, efficient coding to work. However, the contradiction is only apparent: the generative model that is the world is highly complex, even if stable, and unknowable in its entirety. At any

trajectory of the system, the system's trajectories might never actually lie *on* the attractor itself, in contrast to the examples shown in Figures 10-12. As a perhaps intriguing sidenote, we might therefore view the world itself as an attractor of the system corresponding to our agent: the instantaneous state of the system is continuously being drawn both towards the dynamical attractor landscape corresponding to its internal approximated p(o), and towards the dynamical attractor landscape that is the world.

## 5. Action is primary

Now that I have described attractors as a potentially interesting unifying concept underlying the various factors that determine the landscape of approximated p(o), it is worth revisiting the account of action from this perspective. I started this paper by discussing how an agent might optimally encode observations, and how tracking their probabilistic structure provides a useful and energy-efficient means of doing so (section 2). This clearly perceptual starting point was motivated by Bayesian brain theories being primarily dominant in the sensory cognitive (neuro)science. Approximating the probabilistic structure of observations entails minimizing the mismatch between their tracked and actual distributions. This key mechanism of mismatch minimization then also functions to explain action: action is the result of minimizing the mismatch between goal (input) states and actual (input) states (section 3).

In fact, this perspective on action generation does not require that we think about the probabilistic structure of perception at all. We can put aside the interpretation of the attractor landscape as reflecting probability mass, and what remains is something like the fairly trivial: to achieve desired states, one must act. The crucial and perhaps less trivial insight is that such desired states are always formulated over *observations*: agents strive to observe X, and therefore act to bring their observations in line with X (section 3.2). A core principle here is that the reduction of the discrepancy between some pattern of input (which we can interpret as either a goal, or *a priori* expected, or necessary, or desired, …) and actually observed input is a general mechanism underlying much of organism functioning, including perception. In other words: in the space of states describing (input to) organisms there always exist attractors.

The simplest interpretation we can give to these attractors is as goal states. Even extremely simple agents are amenable to such interpretation. Consider, for example, the Braitenberg (Braitenberg, 1986) vehicle depicted in Figure 14A. It has two light sensors and two wheels, and when light falls on the left light sensor, the right wheel starts spinning faster, and vice versa. The right wheel spinning faster than the left causes the vehicle to turn left (and, again, vice versa). Therefore, this wiring diagram causes the vehicle to always "chase" light sources; it will always turn towards a (possibly moving) light source and approach it. There exists an attractor state of this vehicle; namely, the state in which it is turned straight towards the light. Or, phrased in the dimensions in which the vehicle senses the world: equal light striking both given moment, an agent can therefore never perfectly predict any incoming observations, and therefore perturbations remain.

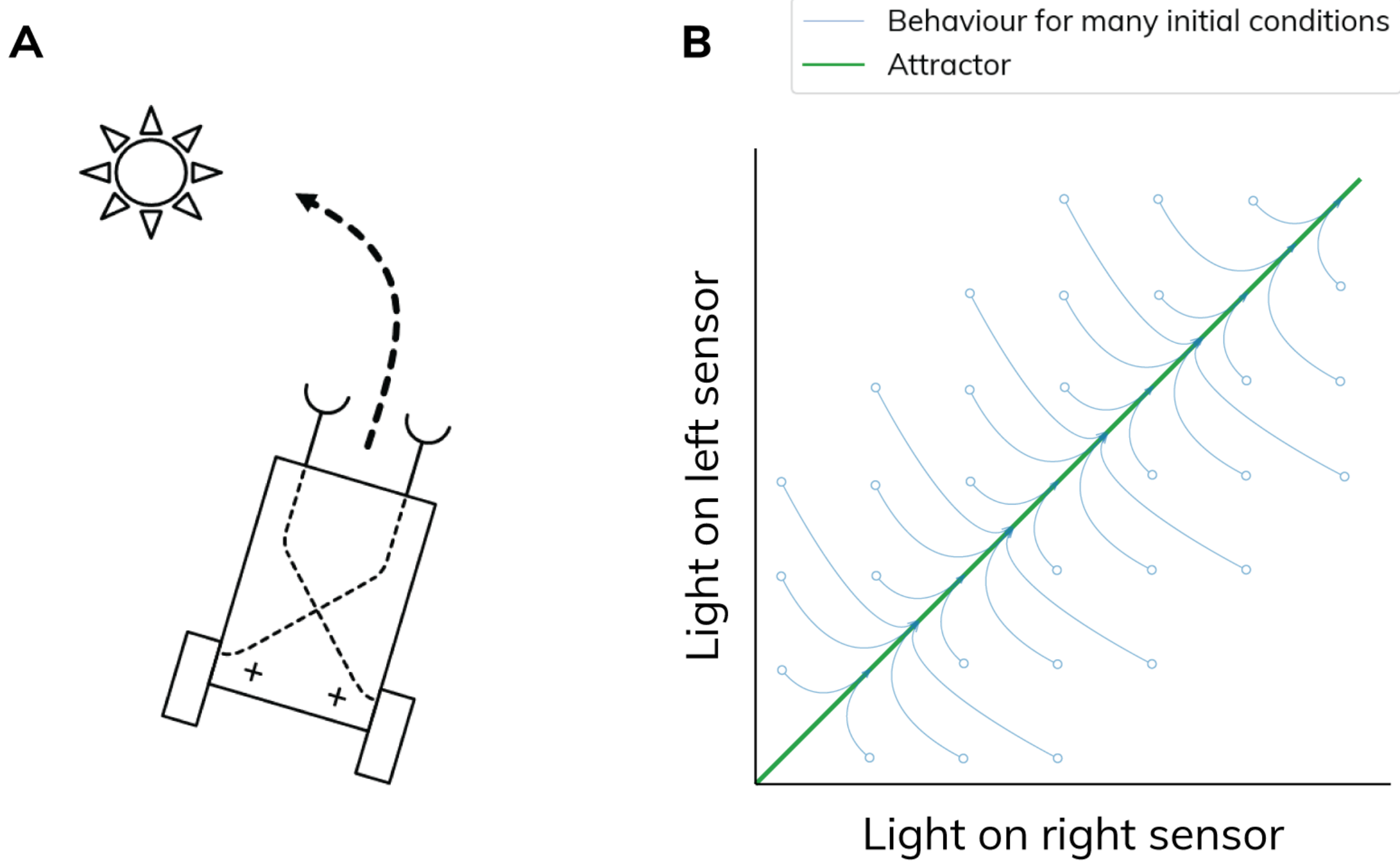

**Figure 14. (A)** A Braitenberg vehicle with two light sensors that are crossed and positively coupled to two rear wheels. It displays phototaxis, i.e. it chases light sources. **(B)** The behaviour of this simple agent can be characterized by the structure of its two-dimensional input state. The situation where equal light falls on both sensors is an attractor of this system. (Sidenote: If the vehicle is given free reign and a stationary external light source, a fixed point in the top right region of input space is actually the system's only attractor, corresponding to the situation where the vehicle is touching the light source head-on; not shown.)

sensors is an attractor. Because this attractor exists, the vehicle exhibits behaviour that minimizes mismatch between actual and 'desired' input. The upshot is: any agent that behaves in response to input (and that includes all biological organisms) can be characterized by outlining the attractor structure present in its input space.

We thus see that the existence of attractors in the input space of agents is a general principle, not limited to complex organisms but present even in some of the simplest agents conceivable. In biological systems, this minimization of mismatch through action is likely phylogenetically ancient. In fact, the reason why organisms evolved sense organ(elle)s to begin with is to be able to then influence the world such that they would observe conditions beneficial to their survival.

The sensory world of mammals is vastly richer than that of the Braitenberg vehicle. Over evolutionary timescales, organisms have developed ever richer means of reducing the mismatch between 'attracting' and observed input states, thereby erecting ever more complex attractor landscapes in their input spaces. For complex animals, a particularly fruitful attractor landscape to have is one that tracks the probabilistic structure of observations: doing so results in lower (long-term) mismatch, as outlined in section 2. Such an attractor has clear and desirable regularizing and information-extracting properties. We can now appreciate that the original imperative for striving to achieve minimal discrepancy between the attracting (now a much broader term than 'goal') region of input space, and the actually observed input, is phylogenetically preserved. While the Bayesian perspective on brain function primarily features in the *sensory* cognitive (neuro)science, we then see that the principle that prescribes the underlying machinery has its roots in the generation of *action*.

## 6. Summary and conclusions

I started this paper by outlining how the familiar narrative of the Bayesian brain, attempting to figure out hidden causes of observations, can be generalized by positing that the brain is tracking the probabilistic structure of those observations themselves. This viewpoint led to the idea that prior expectations, learned from the past, exert a regularizing influence over the encoding of the future (section 2). In section 3, I described how these prior factors, pulling agent state towards them, can lead to action. However, once we interpreted action as stemming from 'prior expectations', the force induced by these prior expectations is no longer adequately describable as such; we ran into the limits of the probability metaphor. In section 4, to resolve this issue, I explained how both perception-regularizing prior expectations and behaviour-inducing goals can be understood as attractors in the input state of agents. Finally, in section 5, I argued that the existence of such attractors is a universal phenomenon present in all agents that sense their environment and act upon it. The often-demonstrated Bayesian nature of our perceptual systems is one special case of this phenomenon; a particularly valuable one, since it enables the extraction of useful information from the world.

## 7. Acknowledgements

I am grateful to Floris de Lange for providing helpful comments on an earlier draft, and to Jelle Bruineberg and a second anonymous reviewer for further helpful comments. This work was supported by The Netherlands Organisation for Scientific Research (NWO Veni grant 016.Veni.198.065).